\documentclass[twocolumn,showpacs,preprintnumbers,amsmath,amssymb,
superscriptaddress]{revtex4}
\usepackage{graphicx}% Include figure files
\usepackage{dcolumn}% Align table columns on decimal point
\usepackage{bm}% bold math

\begin{document}

\preprint{}

\title{Neutron Scattering Study of URu$_{2-x}$Re$_x$Si$_2$ with $x$ =
  0.10: Driving Order towards Quantum Criticality}

\author{T.~J.~Williams}
 \affiliation{Department of Physics and Astronomy,
   McMaster University, 1280 Main St. W.,
   Hamilton, ON, Canada, L8S 4M1}

\author{Z.~Yamani}
 \affiliation{National Research Council, Canadian Neutron Beam Centre,
   Chalk River Laboratories,
   Chalk River, ON, Canada, K0J 1J0}

\author{N.P.~Butch}
 \affiliation{Department of Physics,
   University of California San Diego,
   La Jolla, CA, USA, 92093}

\author{G.M. Luke}
\affiliation{Department of Physics and Astronomy,
   McMaster University, 1280 Main St. W.,
   Hamilton, ON, Canada, L8S 4M1}
\affiliation{Canadian Institute for Advanced Research, Toronto, 
Ontario, Canada, M5G 1Z8}

\author{M.B.~Maple}
 \affiliation{Department of Physics,
   University of California San Diego,
   La Jolla, CA, USA, 92093}

\author{W.J.L.~Buyers}
 \altaffiliation{author to whom correspondences should be addressed:
   E-mail:[william.buyers@nrc.gc.ca]}
 \affiliation{National Research Council, Canadian Neutron Beam Centre,
   Chalk River Laboratories,
   Chalk River, ON, Canada, K0J 1J0}
 \affiliation{Canadian Institute for Advanced Research, Toronto, 
Ontario, Canada, M5G 1Z8}

\date{\today}

\begin{abstract}
We report inelastic neutron scattering measurements in the hidden
order state of URu$_{2-x}$Re$_x$Si$_2$ with $x$ = 0.10.  We observe that
towards the ferromagnetic quantum critical point induced by the negative 
chemical pressure of Re-doping, the gapped incommensurate fluctuations are 
robust and comparable in intensity to the parent material.  As the Re 
doping moves the system toward the quantum critical point, the commensurate 
spin fluctuations related to hidden order weaken, display a shortened 
lifetime and slow down.  Halfway to the quantum critical point, the hidden 
order phase survives, albeit weakened, in contrast to its destruction by 
hydrostatic pressure and by positive chemical pressure from Rh-doping.
\end{abstract}

%\keywords{}

\maketitle

In the field of strongly correlated heavy fermion systems, one of the most
puzzling long-standing issues is the nature of the hidden order
below the large specific heat jump at T$_0$ = 17~K in
URu$_2$Si$_2$~\cite{Palstra_85,Maple_86,Schlabitz_86,Broholm_87}.  A 
superconducting phase also follows below 1.2 K.  Despite much research over 
the past 25 years~\cite{Bonn_88,Buyers_94,Mason_95,Bourdarot_05, Niklowitz_11} 
the order parameter remains unknown.  The small antiferromagnetic moment of 
0.03~$\mu_B$ that develops below 17 K cannot explain the large specific heat 
jump at this second-order phase transition.  Antiferromagnetism therefore
cannot be the main cause of the hidden order.  The system appears 
to have condensed into a new phase of matter for which the order parameter
and associated symmetries differ from conventional expectations.  In
previous work, we eliminated crystal fields and orbital currents as
a source of hidden order~\cite{Wiebe_04,Wiebe_07}.  The
spins fluctuating above the 17 K transition are centred on the incommensurate 
wavevector (1 $\pm$~$\delta$, 0, 0) with $\delta$~=~0.4. Emerging from that 
wavevector is a high-velocity cone of strongly damped gapless excitations 
that extend over a finite region of the Brillouin zone.  They provide 
evidence of itinerant, rather than localized spins.  In the precursor
phase to hidden order, Wiebe {\it et al}. calculated that these gapless
spin fluctuations give rise to a term in the specific heat that
closely accounts for the magnitude of the giant specific heat linear
in T that was previously attributed entirely to electrons~\cite{Wiebe_07}.
In addition, the decrease of the specific heat below T$_0$ is now
understood~\cite{Wiebe_07} to arise from the formation of a spin gap below 
T$_0$.  More recently, the excitations have been interpreted as the response 
of itinerant spins to Fermi surface nesting, similar to that of chromium, 
and specific nesting vectors were proposed~\cite{Janik_09}.  Evidence of Kondo 
hybridization arises from STM studies~\cite{Schmidt_10}.  Despite this 
considerable progress, the symmetry of the order parameter that condenses in 
the hidden order phase remains unknown.

\begin{figure}[htb]
\includegraphics[angle=0,width=\columnwidth]{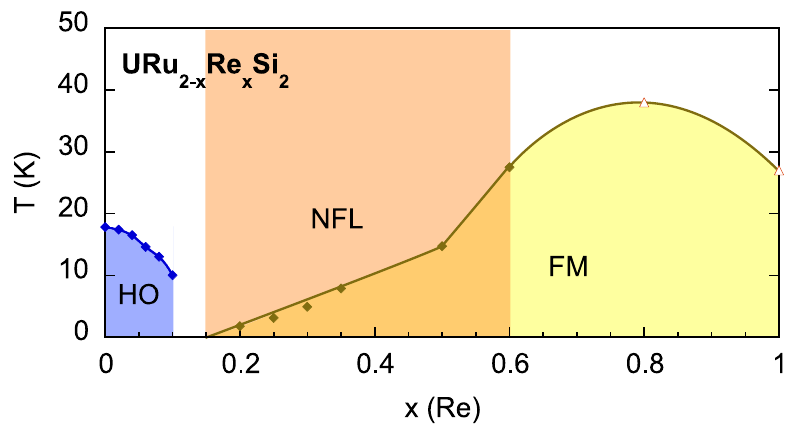}
\caption{\label{phase_diagram}
Magnetic phase diagram of URu$_{2-x}$Re$_x$Si$_2$ showing the
antiferromagnetic (hidden order) and ferromagnetic
phases.  From Ref.~\cite{Butch_10}, also see~\cite{Butch_09, Bauer_05}.}
\end{figure}

Another route to discovering the hidden order symmetry is to move
away from the object of interest by applying hydrostatic pressure, or chemical
pressure by doping.  Hydrostatic pressure has been recently found to
cause commensurate condensation of a larger antiferromagnetic moment
of 0.3~$\mu_B$ and the collapse of the strong commensurate spin
excitations~\cite{Aoki_09, Villaume_08}.  The hidden order phase is eventually 
lost with application of pressure beyond about 1.5 GPa.  On the
contrary, the hidden order phase can be retained by applying
negative chemical pressure.  This is done by expanding the lattice
constants by alloying with rhenium as discovered in the pioneering
work of Dalichaouch {\it et al.}~\cite{Dalichaouch_89, Dalichaouch_90}. However,
the Re replacement of Ru atoms in URu$_{2-x}$Re$_x$Si$_2$ reduces the hidden 
order transition from 17~K to 13~K for $x$~=~0.10~\cite{Butch_10, Butch_09, 
Bauer_05}.  It also reduces the
superconducting transition from 1.5~K to 0.23~K at $x$~=~0.01.
Interestingly, this represents a first step on the way to reaching a
quantum critical point where the hidden order gives over to
ferromagnetism~\cite{Butch_10}.  The system exhibits non-Fermi liquid behavior 
for $0.15 < x < 0.6$. The magnetic phase diagram is shown in
Fig.~\ref{phase_diagram}.

We performed neutron scattering measurements on the DUALSPEC
spectrometer at the C5 beamline of the NRU reactor at Chalk River
Laboratories.  Single crystals of URu$_{1.9}$Re$_{0.1}$Si$_2$ and
URu$_2$Si$_2$ were aligned in the (H 0 L) scattering plane.  Unless
otherwise stated, the experiments were performed with a setup and
collimation of 0.53$^{\circ}$-PG-0.55$^{\circ}$-S-0.85$^{\circ}$-PG-1.2$^{\circ}$
with a final scattering energy $E_f$~=~14.6~meV, using two pyrolytic graphite 
(PG) filters, a vertically focusing PG monochromator and a flat PG analyzer. 
The fast neutron contribution to the background was measured by rotating the 
analyzer from the Bragg reflection by five degrees.

We find that there is no elastic incommensurate scattering at 2~K at the
wavevector (1.4~0~0) nor at the commensurate antiferromagnetic point (1~0~0).
The latter is in contrast to the parent material, where the weak elastic
scattering at the commensurate point is believed to arise from 
a minority phase~\cite{Wiebe_04}.  

\begin{figure}[htb]
\includegraphics[angle=0,width=\columnwidth]{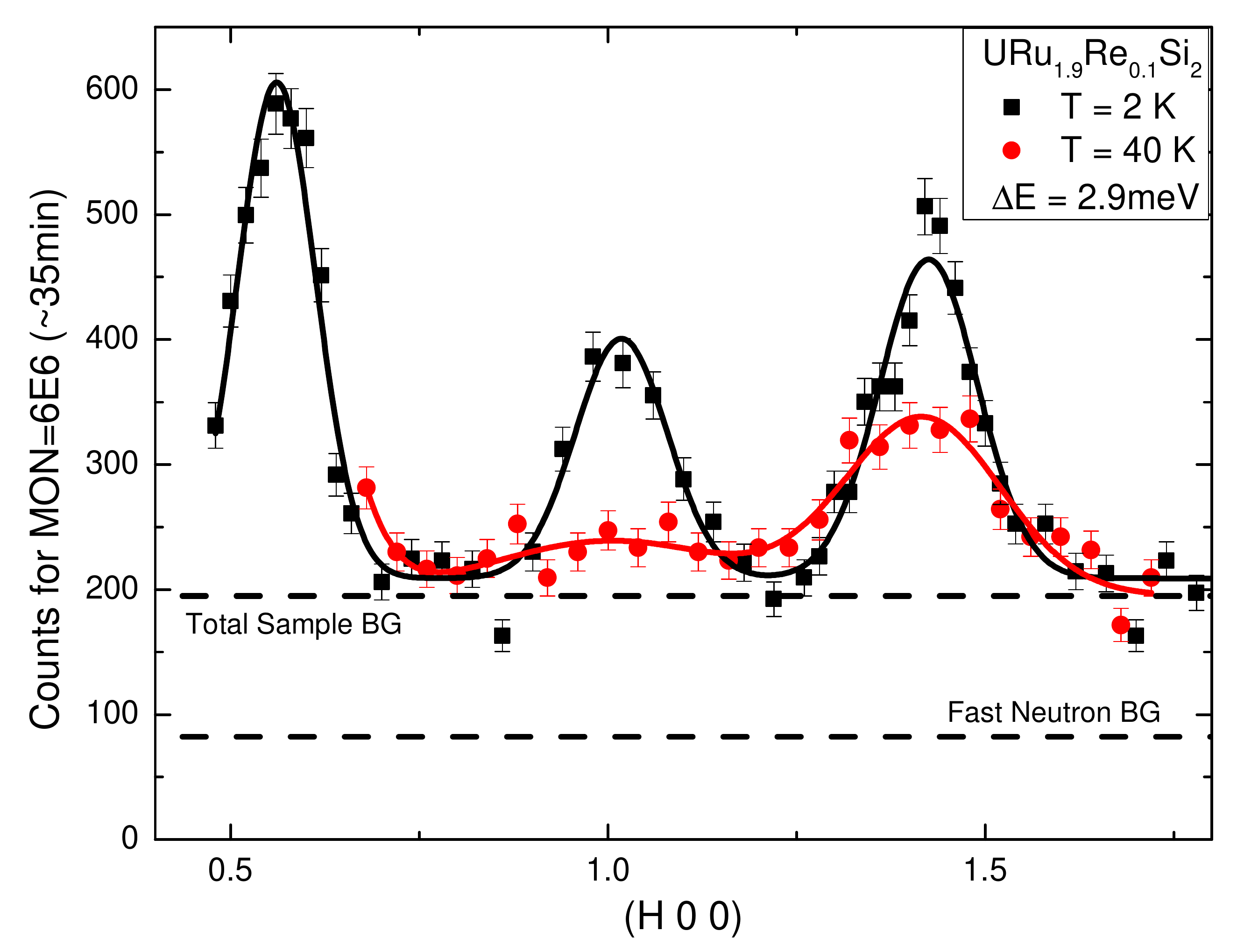}
\caption{\label{H00_Tdep}
Inelastic scattering along the (H~0~0) direction for an energy transfer
of 2.9~meV at T~=~2~K (black squares) and T~=~40~K (red circles).
At 40~K well above the hidden order transition at 17~K, there is partial 
suppression of the incommensurate fluctuations. In contrast, there is an 
almost complete suppression of the commensurate fluctuations associated 
with the hidden order. The lines are Gaussian fits at each of the three 
peak positions.}
\end{figure}

Fig.~\ref{H00_Tdep} shows the dependence of scattering along the (H00) 
direction at an energy transfer of 2.9~meV at 2~K and 40~K. This measurement 
represents the raw data.  At 2~K, the scattering shows the relative strength 
expected from the magnetic form factor at equivalent incommensurate 
wavevectors (0.6~0~0) and (1.4~0~0). At this temperature, the scattering at 
the commensurate position (1~0~0) also exhibits comparable strength.  The 
fast neutron background is shown, as well as the total sample background, 
obtained from the fitting, described later in the text.

\begin{figure}[htb]
\includegraphics[angle=0,width=\columnwidth]{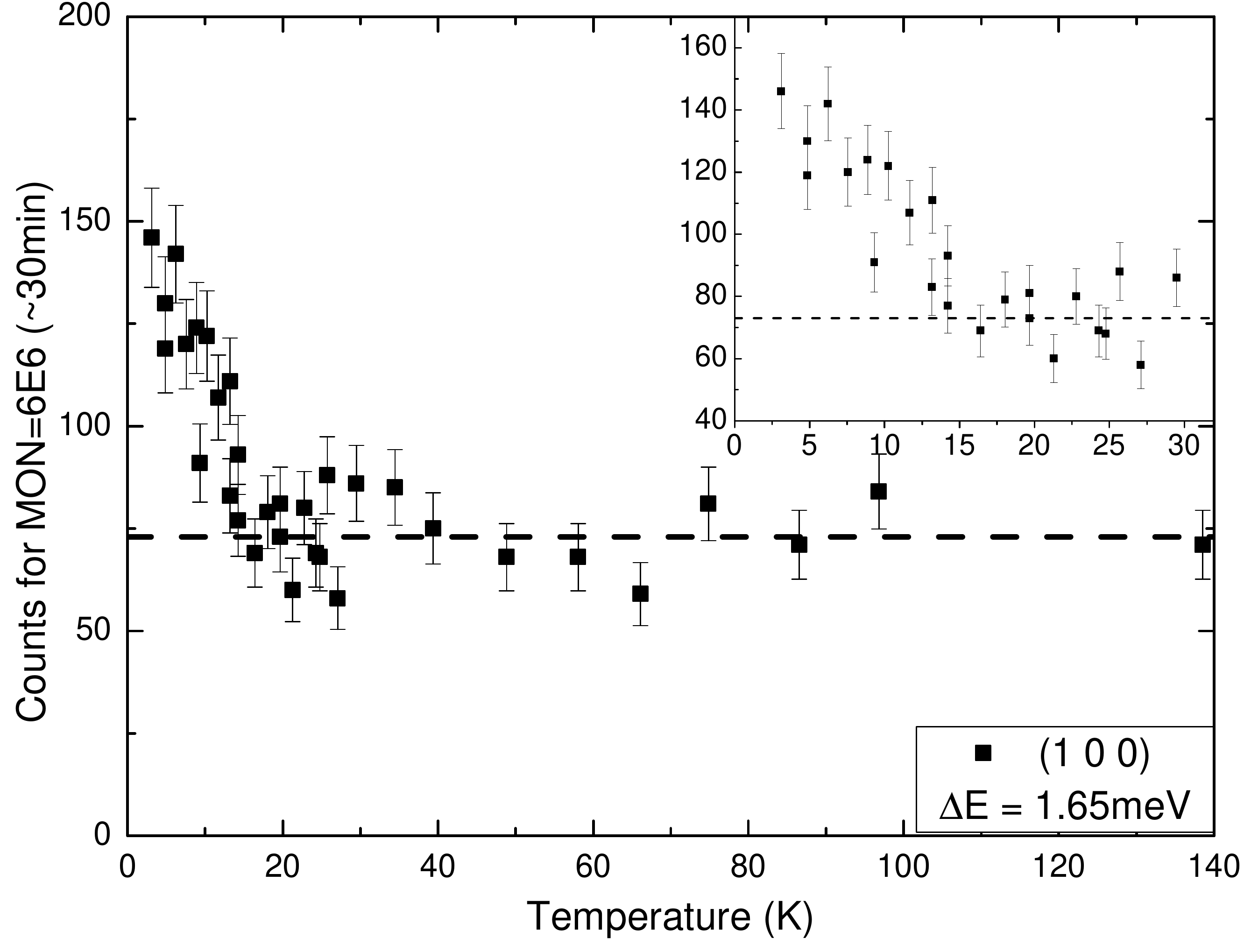}
\caption{\label{100_Tdep}
The temperature dependence of the commensurate (1~0~0) fluctuations in
URu$_{1.9}$Re$_{0.1}$Si$_2$ at 1.65~meV.  A discontinuity in slope is observed 
at around the onset of the hidden order phase.  The inset shows a close-up of
the data below 30~K.}
\end{figure}

The change in scattering as a function of temperature, both at the 
commensurate and incommensurate positions, can be used to identify the 
hidden order transition. At 40~K, well above the hidden order transition, 
Fig.~\ref{H00_Tdep} shows that the incommensurate scattering remains 
strong, with roughly half the intensity, but the commensurate fluctuations 
decrease substantially.  This is consistent with the suggestion that 
commensurate (1~0~0) dynamic spin excitations are a signature of the hidden 
order phase~\cite{Bourdarot_10}, whereas the incommensurate scattering is 
present in both the paramagnetic and hidden order phases.  
Figure~\ref{100_Tdep} shows the temperature dependence of the commensurate 
(1~0~0) fluctuations at 1.65~meV.  This measurement was performed under 
different experimental conditions, which accounts for the change in the 
background compared to Fig.~\ref{H00_Tdep}, but with the same array of 
single crystals.  A discontinuity in the temperature dependence of the peak 
is observed around $\sim$13 K, which may indicate the onset of the hidden 
order phase.  This change is not as clear as that observed in the parent 
material~\cite{Bourdarot_10}, likely due to electronic disorder associated 
with the Re substitution.  However, the reduction in this dynamic measure 
of T$_0$ is consistent with references 
\cite{Dalichaouch_89,Dalichaouch_90,Butch_10}.

\begin{figure}
\includegraphics[angle=0,width=\columnwidth]{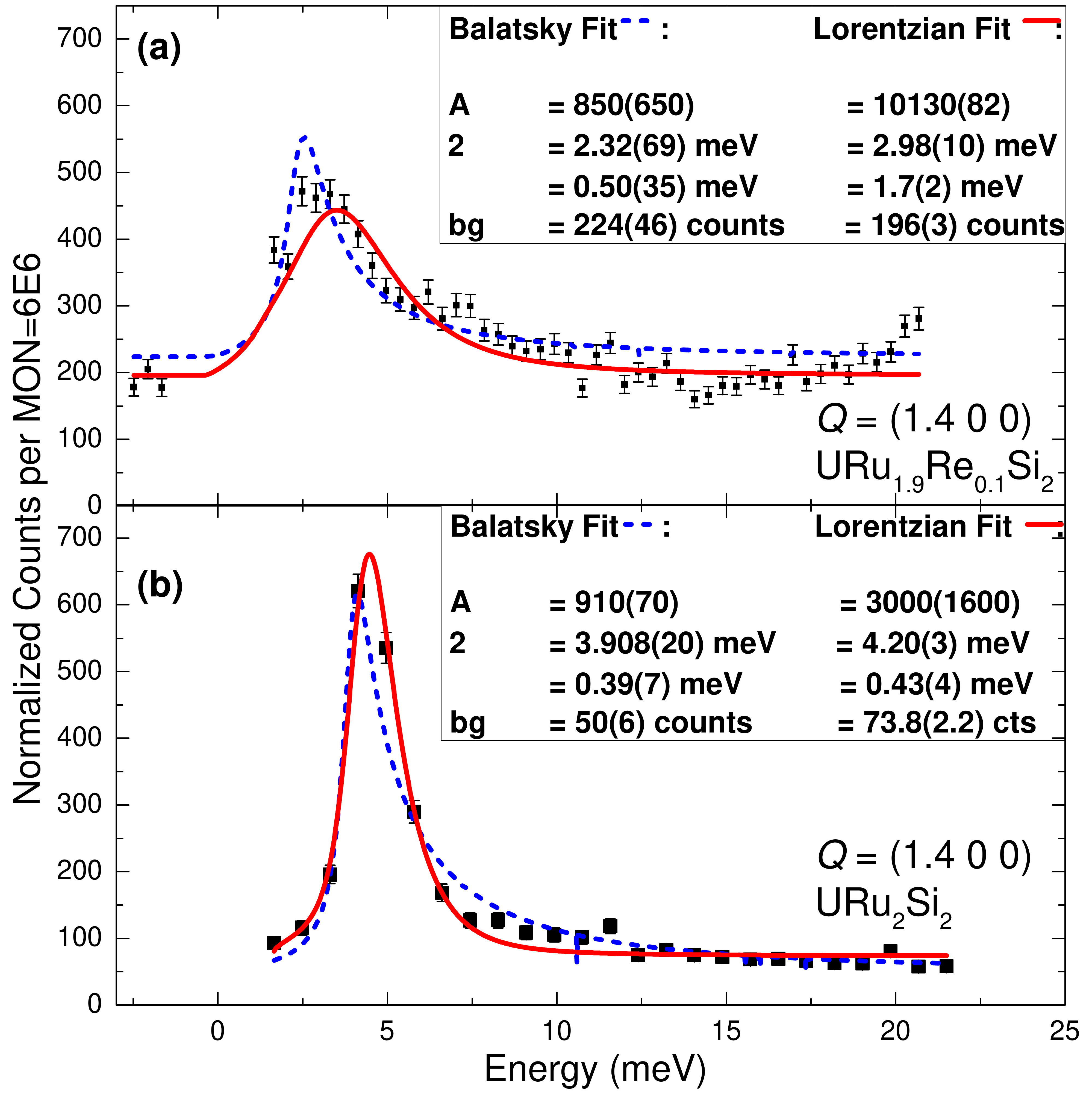}
\caption{\label{1400_Balatsky}
(color online) The fits to the incommensurate fluctuations (a) for
URu$_{1.9}$Re$_{0.1}$Si$_2$ at T~=~2~K and (b) for pure URu$_2$Si$_2$ at 
T~=~3~K and 5~K (combined data).  The blue line is the fit to Eq.~\ref{eq1} 
and the red line is a fit to a Lorentzian of energy $2\Delta$, amplitude 
$A$ and damping $2\gamma$=FWHM, convoluted with the resolution function, as 
described in the text.  For the Re-doping, the nesting gap energy, $\Delta$, 
is reduced to 60\% of its value in the pure system.}
\end{figure}

Energy scans at Q = (1.4 0 0), comparing the 10$\%$ Re-doped with the pure
crystal are shown in Fig.~\ref{1400_Balatsky}.  The data have been
normalized to constant volume for the two crystals via phonon measurements
at (2.3 0 0) and (1.8 0 0) respectively.  The spectrum of
URu$_{1.9}$Re$_{0.1}$Si$_2$ also exhibits an incommensurate spin gap similar 
to that in pure URu$_2$Si$_2$~\cite{Janik_09,Broholm_91}. However the gap 
value is lowered by doping.  

\begin{figure}
\includegraphics[angle=0,width=\columnwidth]{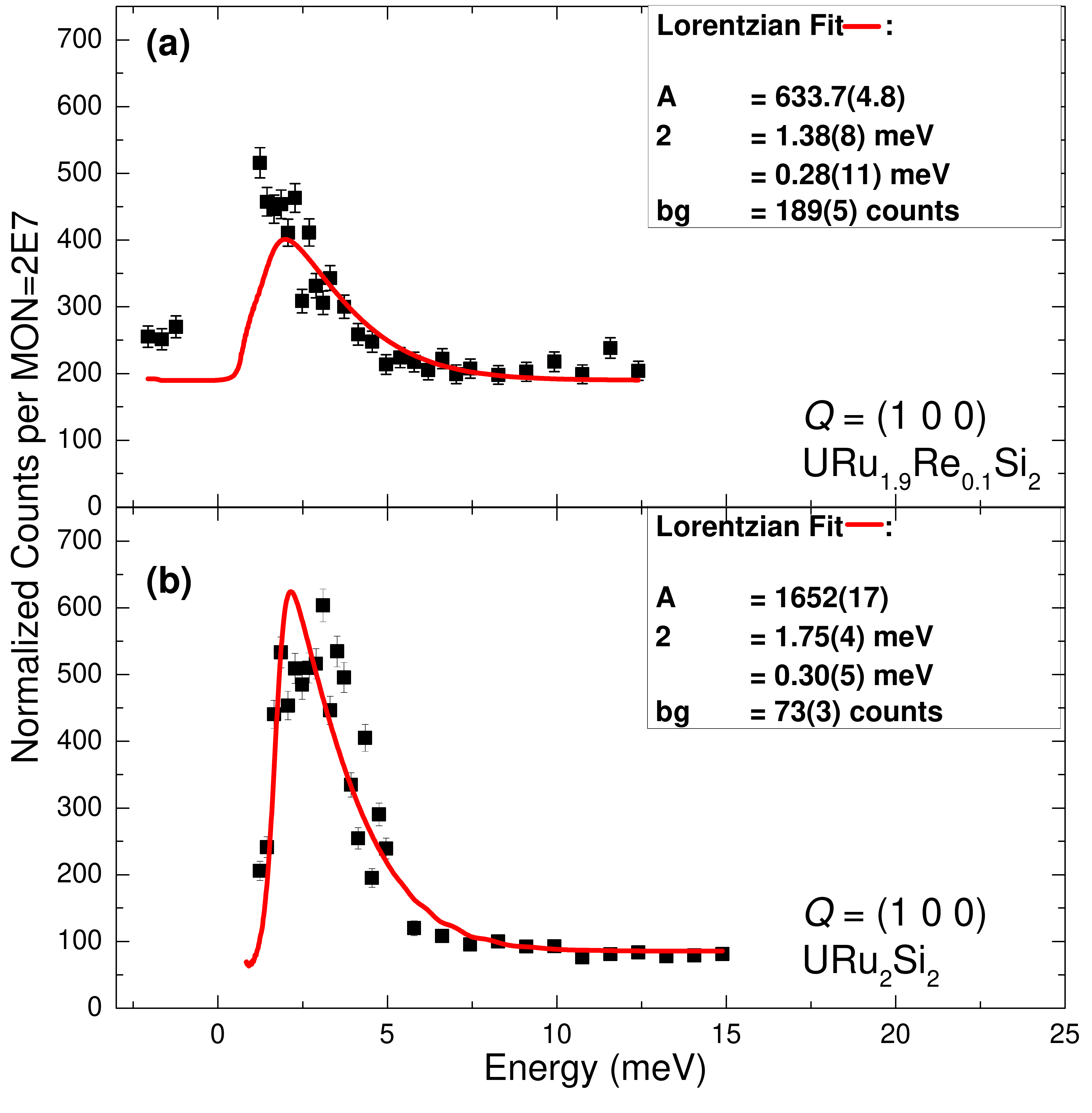}
\caption{\label{all_fits}
(color online) (a) The inelastic scattering at (1~0~0) in the 
URu$_{1.9}$Re$_{0.1}$Si$_2$ sample, measured at T~=~2~K. (b) The (1~0~0) 
inelastic scattering in the pure URu$_2$Si$_2$system, measured at T~=~3~K 
and 5~K (combined data).  Both were fit to a Gaussian for the elastic 
scattering, and a Lorentzian of energy $\omega_{\vec{Q}^{\ast}}=2\Delta$, 
amplitude $A$ and damping $2\gamma$~=~FWHM for the inelastic peak, convoluted 
with the resolution function.  The commensurate fluctuations of the hidden 
order phase are damped in the prescence of Re-doping.}
\end{figure}

The normalized intensity comparison of Fig.~\ref{1400_Balatsky} shows doping 
has reduced the intensity at the incommensurate wavevector by a factor of 2 
(obtained from the integrated intensity of the peaks).  Re-doping also 
increases the spectral width as seen in Fig.~\ref{1400_Balatsky}, showing 
that the fluctuations are highly damped by doping.  The slowing of 
fluctuations is more dramatic at the commensurate hidden order wavevector, 
as shown in Fig.~\ref{all_fits}.  There, the lifetime is so short that the 
characteristic energy barely gives a peak in the spectrum. It may also 
signify the destruction of perfect nesting by charge impurities.

The data in Fig.~\ref{1400_Balatsky} are well-described by the theoretical
model of Balatsky~\cite{Balatsky_09}, given by:

\begin{equation}
\chi^{zz}(\vec{Q}^{\ast},\omega)=
A^2 \lvert \Delta_{\vec{Q}^{\ast}} \rvert^2 \int
\frac{1}{\sqrt{E^2-\Delta_{\vec{Q}^{\ast}}^2}} \frac{1}{\omega^2-4E^2} dE
\label{eq1}
\end{equation}

to which a constant background (bg) has been added, shown by the blue lines 
in Fig.~\ref{1400_Balatsky}. The theory is based on 
a spin resonance in the hidden order state, involving transitions between 
nested parts of the Fermi surface, separated by $\vec{Q}^{\ast}$~=~(1.4~0~0).  
This spin resonance is caused by partial Fermi surface nesting which leads 
to the appearance of a particle-hole condensate~\cite{Balatsky_09}.  This 
nesting can occur for a continuous range of wavevectors, from (1~0~0) to the 
peak at (1.4~0~0), corresponding to the maximum nesting vector between the 
two Fermi surface pockets proposed from band structure 
calculations~\cite{Janik_09,Balatsky_09}. The fermion energies are assumed 
to rise quadratically above a hidden order gap $\Delta_{\vec{Q}^{\ast}}$, 
allowing pairs of excitations to contribute to the dynamic susceptibility 
measured by neutron scattering~\cite{Balatsky_09}.  Thus the gap in the spin 
spectrum in Fig.~\ref{1400_Balatsky} is equal to $2\Delta_{\vec{Q}^{\ast}}$.

In Fig.~\ref{1400_Balatsky}, the models for the spectrum have been 
convoluted with the 4D instrumental resolution function using 
Reslib~\cite{Zheludev_07}.  The required spin velocities in three directions 
were taken from~\cite{Broholm_91}.  The spectrum was broadened slightly to 
deal with the square root singularity of the Balatsky nesting Ansatz of 
Eq.~\ref{eq1}, by adding a small imaginary part $\gamma$ to the frequency in 
such a way that it is analogous to the broadening $\gamma$ of 
Fig.~\ref{1400_Balatsky} described in relation to Eq.~\ref{eq2}.  The 
data have been corrected for higher order perturbation of the monitor rate.

Recent measurements of the inelastic scattering along (H~0~0) in the parent 
compound have been analyzed~\cite{Bourdarot_10} in terms of a peak and a 
continuum. In contrast we find that for both the pure and doped systems,
the tail to high-energy is well-described by the asymmetry of the Balatsky 
nesting equation (\ref{eq1}) and no additional continuum is needed.  We 
suggest that the doped Re atoms render the nesting less perfect and so allow 
spin fluctuations over a wider range of energies.  This is consistent with 
the picture of the incommensurate spin resonance that leads to 
Equation~\ref{eq1}.

Attempts to fit the commensurate fluctuations at (1~0~0) to the Balatsky 
theoretical model did not converge.  The data for the
commensurate and incommensurate excitations for both doped and pure
compounds were therefore fitted to Lorentzians for comparison purposes, 
given by:

\begin{equation}
I(\vec{Q}^{\ast},\omega)=
\frac{A}{\omega_{\vec{Q}^{\ast}}} \cdot \left [ 
\frac{1}{(\omega-\omega_{\vec{Q}^{\ast}})^2
+\gamma^2} - \frac{1}{(\omega+\omega_{\vec{Q}^{\ast}})^2
+\gamma^2} \right ]
\label{eq2}
\end{equation}

This was multiplied by a Bose factor, and convoluted with the resolution 
function, as described above.  The commensurate fluctuations at (1 0 0) for 
the Re-doped and parent samples are shown in Fig.~\ref{all_fits}(a) and 
Fig.~\ref{all_fits}(b), respectively.  The resolution conditions for the 
(1 0 0) energy scan (Fig.~\ref{all_fits}(a)) were slightly different, with 
a setting of 0.53$^{\circ}$-PG-0.48$^{\circ}$-S-0.55$^{\circ}$-PG-1.2$^{\circ}$. 
Compared to the parent compound with a spin gap of 1.75~meV, the commensurate 
fluctuations are peaked at a lower energy of 1.38~meV, a reduction of 72\%, 
which tracks the reduction of T$_0$. Within a nesting 
picture, it appears that the Re impurities (chemical pressure) greatly 
weaken the nesting present in the pure system.

As temperature is increased, the commensurate fluctuations are destroyed much 
more quickly than are the incommensurate fluctuations at (1.4 0 0). Thus the 
hidden order gap for (1 0 0) antiferromagnetic fluctuations becomes less well 
defined.  Both the commensurate (Fig.~\ref{all_fits}(a)) and incommensurate
(Fig.~\ref{1400_Balatsky}(a)) spectral form is that of a resonant frequency that
decays into the Re-induced continuum of the itinerant particle-hole states.
The Re doping achieves this, we suggest, by $\vec{Q}$-broadening of the
nesting that gave the well-defined spectral onset above the gap in the pure
hidden order system URu$_2$Si$_2$.  The relatively large spin 
velocities~\cite{Broholm_91} then convert the $\vec{Q}$-broadening into the 
observed spectral broadening.  Our results suggest that the gap may vanish 
when the quantum phase transition to ferromagnetism is reached.

\begin{figure}[thb]
\includegraphics[angle=0,width=\columnwidth]{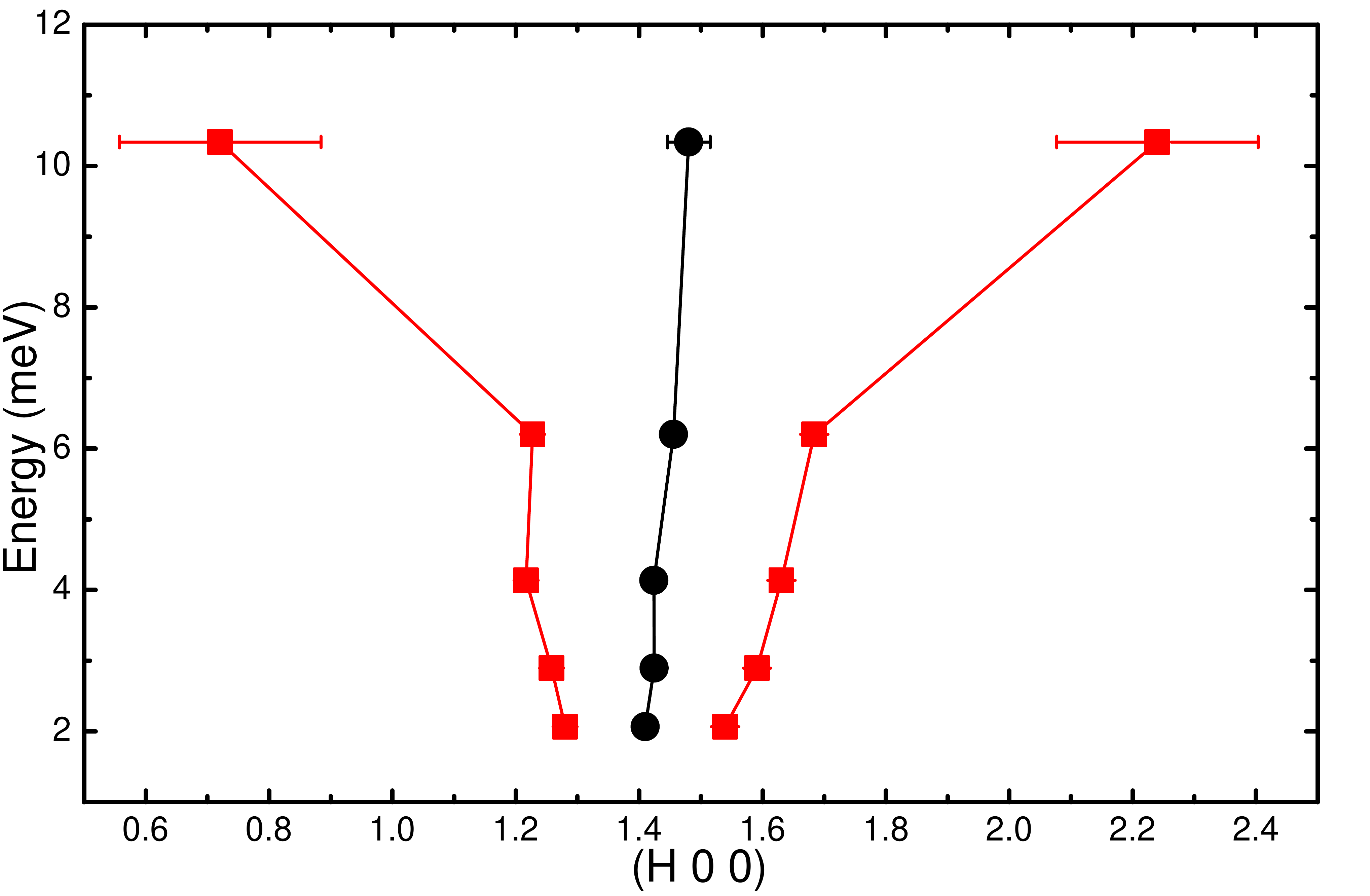}
\caption{\label{dispersion}
Locus of the half widths in reduced wavevector H of spin fluctuations
around (1.4 0 0).  These inverse dynamic correlation
lengths are narrow in wavevector for low energies, but broaden
dramatically at higher energies.  The black vertical line shows the
centroid of the scattering arising from a column of itinerant excitations
in the E-$\vec{Q}$ plane.  The error bars for low energy fits lie within 
the symbol size.}
\end{figure}

We also studied the width in wavevector of the column of spin
fluctuations that emerges from the incommensurate
points.  Constant-energy scans were measured along (H 0 0) at energies 
ranging from 2.1~meV to 10.3~meV as shown in Fig.~\ref{dispersion}.  These 
$\vec{Q}$-$E$ patterns with Re present are very similar to those observed 
in the pure material~\cite{Janik_09}.  The FWHM in H at 10~meV is anomalously 
large because of overlap with spin cones emanating from (1 0 0) and phonons 
from (2 0 0), but the low-energy correlation width is accurate.  We note that 
the intensity and peak energy of the excitations have decreased from the parent
material, but their $\vec{Q}$-width and hence dynamic spin correlation lengths
remains unchanged.  Thus those incommensurate fluctuations that are not a
primary signature of hidden order like those around (1.4 0 0) are little
affected by doping or by temperature.  Thereby, the predominant behaviour that
survives the approach to the quantum critical point is the robust cone of
gapped incommensurate fluctuations, similar to the pure material.

Our results demonstrate that the commensurate spin fluctuations lose much
of their collective peaking as nesting is disturbed by approach to the
presumed ferromagnetic~\cite{Bauer_05} quantum critical point.  In addition, 
we find that the lifetime of the spin fluctuations, or more likely the 
fermions from which they arise, are shorter in the Re-doped compound.  The 
observation of a commensurate spin gap indicates that the hidden order phase 
survives at least half-way to the quantum critical point albeit in a weakened 
form.  Our neutron results therefore show that the effect of Re doping, in 
contrast to the antiferromagnetic enhancement and hidden order destruction by 
Rh-doping\cite{Butch_09,Butch_10}, is to weaken, but surprisingly, not to 
destroy the hidden order on approach to the quantum phase transition 
to ferromagnetism.  This is consistent with the part of the hidden order phase 
boundary inferred from specific heat measurements~\cite{Butch_09,Bauer_05}.

\begin{acknowledgments}
We appreciate the hospitality and support of the NRC Canadian
Neutron Beam Center at Chalk River laboratories and the technical support of 
R. Sammon. Research at McMaster University is supported by NSERC. 
Research at UCSD was supported by the U.S. Department of Energy under Grant 
Number DE-FG02-04ER46105.
\end{acknowledgments}

\end{document}